\documentclass[twocolumn,showpacs,preprintnumbers,prl,amsmath,amssymb]{revtex4}

\usepackage[pdftex]{graphicx}
\usepackage{dcolumn}
\usepackage{bm}
\begin{document}
\title{Magnetic structures and the Ce-Fe coupling induced Fe spin reorientation in CeFeAsO single crystal}
\author{Qiang Zhang$^{1,2}$}\email{qzhangemail@gmail.com} 
\author {Wei Tian,$^3$ Haifeng Li,$^{1,2}$ Jong-Woo Kim,$^{4}$ Jiaqiang Yan,$^{1,3,6}$, R. William McCallum,$^{1,5}$ Thomas A. Lograsso,$^{1,5}$ Jerel L. Zarestky,$^1$ Sergey L. Bud'ko,$^{1,2}$ Robert J. McQueeney$^{1,2}$}
\author{David Vaknin$^{1,2}$} \email{vaknin@ameslab.gov}
\affiliation{$^1$Ames Laboratory, Ames, Iowa, 50011, USA \\
$^2$Department of Physics and Astronomy, Iowa State University, Ames, Iowa 50011, USA \\
$^3$Oak Ridge National Laboratory, Oak Ridge, Tennessee 37831, USA\\
$^4$Advanced Photon Source, Argonne National Laboratory, Argonne, Illinois 60439, USA\\
$^5$Division of Materials Sciences and Engineering, Iowa State University, Ames, Iowa 50011, USA\\
$^6$Department of Materials Science and Engineering, University of Tennessee, Knoxville, Tennessee 37996, USA}
\date{\today}

\begin{abstract}
     Neutron and synchrotron resonant X-ray magnetic scattering (RXMS) complemented by heat capacity and resistivity measurements reveal the evolution of the magnetic structures of Fe and Ce sublattices in single crystal CeFeAsO.  The RXMS of magnetic reflections at the Ce $L_{\rm II}$-edge shows
 a magnetic transition that is specific to the Ce antiferromagnetic long-range ordering at $T_\texttt{Ce}\approx$ 4 K with short-range Ce ordering above $T_\texttt{Ce}$,
 whereas neutron diffraction measurements of a few magnetic reflections indicate a transition at $T^{*}\approx$ 12 K with unusual order parameter. Detailed order parameter measurements on several magnetic reflections by neutrons show a weak anomaly at 4 K which we associate with the Ce ordering.
 The successive transitions at $T_\texttt{Ce}$ and $T^{*}$ can also be clearly identified by two anomalies in heat capacity and resistivity measurements. 
The higher transition temperature at $T^{*}\approx$ 12 K is mainly ascribed to Fe spin reorientation transition, below which Fe spins rotate uniformly and gradually in the \textit{ab} plane. 
The Fe spin reorientation transition and short-range Ce ordering above $T_\texttt{Ce}$ reflect the strong Fe-Ce couplings prior to long-range ordering of the Ce. 
The evolution of the intricate magnetic structures in CeFeAsO going through $T^{*}$ and $T_\texttt{Ce}$ is proposed.   
\end{abstract}
 \maketitle

\section{Introduction}
  Since the discovery of high-temperature superconductivity in the fluorine-doped LaFeAsO \cite{Kamihara2008}, the layered iron 
pnictides have attracted considerable attention\cite{Dai2012,Fujitsu2012}. In these systems, superconductivity emerges by doping or by the application of pressure\cite{Fujitsu2012,Zhang2013}. Although doping suppresses both the Fe magnetic order and lattice distortion in the parent compounds, magnetism 
is believed to play a major role in the superconducting pairing mechanism as demonstrated, e.g., by the appearance of a magnetic resonance in the superconducting state.\cite{Christianson2008,Lumsden2009} Therefore,
understanding the magnetism in the iron-based parent compounds can lead to insight about the mechanism that induces superconductivity by doping or pressure.\cite{Zhao2009,Tian2010} 
     
      CeFeAsO is a typical example of rare-earth-containing $R$FeAsO, with a tetragonal-to-orthorhombic structural transition upon cooling at $T_\texttt{S}\approx 150$ K that is followed by the onset of stripe-like Fe AFM order below $T_\texttt{N}\approx 140$  K.\cite{Zhao2008,Jesche2010} Long-range AFM ordering of Ce below $T_\texttt{Ce}\approx$ 4 K was first reported in powder neutron diffraction studies proposing 
 a Ce magnetic structure that required further confirmation.\cite{Zhao2008} This study also suggested that Ce-Fe coupling is relatively strong as it influences the magnetic Fe peaks above $T_\texttt{Ce}$ up to $\approx$ 20 K. Furthermore, in support of strong Fe-Ce coupling, $\mu$SR measurements \cite{Maeter2009} on polycrystalline CeFeAsO show a considerable 3$d$-4$f$ hybridization that leads to a considerable staggered Ce magnetization up to  almost 120 K, far above $T_\texttt{Ce}$. 
However, magnetization, resistivity and heat capacity measurements on polycrystalline\cite{Chen2008} and single-crystalline \cite{Jesche2009} CeFeAsO have been interpreted in terms of
 very weak or no Ce-Fe exchange coupling. 
It is interesting to point out that as an outcome of the Pr-Fe coupling, the iso-structural PrFeAsO
exhibits temperature-induced Fe spin reorientation \cite{Maeter2009,McGuire2009}, whereas for LaFeAsO involving non-magnetic La, the Fe moments are fixed to be along the \textit{a}-axis even at low temperatures \cite{Li2010}.  
Moreover, temperature-induced Fe spin reorientation transition has been observed  in few non-pnictide compounds involving  rare-earth \textit{R} and Fe sublattices, such as \textit{R}FeO$_{3}$\cite{Belov1976}, \textit{ R}Fe$_{2}$ \cite{Belov1976},
 \textit{R}Fe$_{11}$Ti \cite{Yu1995} and \textit{R}$_{2}$Fe$_{14}$B \cite{Pique1996}. In these systems, the competing contributions of Fe and \textit{R} sublattices determine the temperature dependence of Fe magnetic anistropy\cite{Belov1976,Yu1995,Pique1996}. Here, we report a combined magnetic resonant x-ray and neutron scattering study
that is also complemented by heat capacity and resistivity measurements on CeFeAsO single crystal.

\section{Experimental Details}
      The CeFeAsO crystal was grown out of NaAs flux as described previously.\cite{Yan2011} An 8 mg plate-like crystal with a $c$-axis perpendicular to its surface was used for 
all the investigations
 reported here. X-ray powder diffraction of crushed crystals from the same growth batch confirm the $P4/nmm$ space group with lattice constants $a = 3.988$ and
 $c=8.552$ {\AA} (lattice constants at room temperature are also consistent with those obtained from the crystal in this study). 
      
      The resonant x-ray magnetic scattering (RXMS) experiments were performed on a six-circle diffractometer at the 6-ID-B beamline at the Advanced Photon Source (APS, Argonne National 
Laboratory, USA) at and around the Ce $\textit{L}_{II}$
absorption edge $E=6.1642$ keV. The crystal was mounted at the end of a cold finger of a Displex cryogenic refrigerator and oriented so that the scattering plane coincided with 
the (\textit{h}0\textit{l}) crystal plane (orthorhombic notation, note that we keep using the orthorhombic notation for all the structural and magnetic reflections).
 The resonant scattering measurements were carried out with a linearly polarized beam perpendicular to the scattering plane ($\sigma$ polarization). In this geometry, the component of the magnetic moment that is in the scattering plane primarily contributes to the resonant scattering arising from electric dipole transitions from the 2\textit{p}-to-5\textit{d} states, which reflects the symmetry of 
the localized magnetic 4$f$ shell through Coulomb interactions between 4$f$ and 5$d$ bands.\cite{McMorrow2003,Kim2005}. The linear polarization of the scattered radiation for dipole resonant scattering is parallel to the scattering plane ($\pi$ polarization). In contrast,
charge scattering does not change the polarization of the
scattered photons ($\sigma-\sigma$ scattering). \cite{Kim2005,Nandi2009}

      The elastic neutron-scattering measurements were carried out on the HB1A fixed-incident-energy ($E=14.7$ meV) spectrometer using a double pyrolytic
 graphite monochromator (located at the High Flux Isotope Reactor, HFIR, at Oak Ridge National Laboratory, USA). 
For the neutron diffraction, the single crystal was wrapped in aluminum foil and sealed in a helium-filled aluminum can which was then loaded on the cold tip of a closed cycle 
refrigerator. Two series of experiments with (\textit{h}0\textit{l}) and (\textit{h}\textit{k}0) as the scattering planes were performed to collect nuclear and magnetic Bragg reflections. 
Rocking scans were resolution limited, confirming the high-quality of the sample.
 
     Heat capacity measurements on this and other two pieces of crystals (see text) were carried out on a Physical Properties Measurement System (PPMS, Quantum Design) using a semiadiabatic relaxation method.
Resistivity measurements were also performed on PPMS, using a standard four-probe method.

\section{Results and Discussion}    
\subsection{Iron magnetic ordering}
\begin{figure} \centering \includegraphics [width = 1\linewidth] {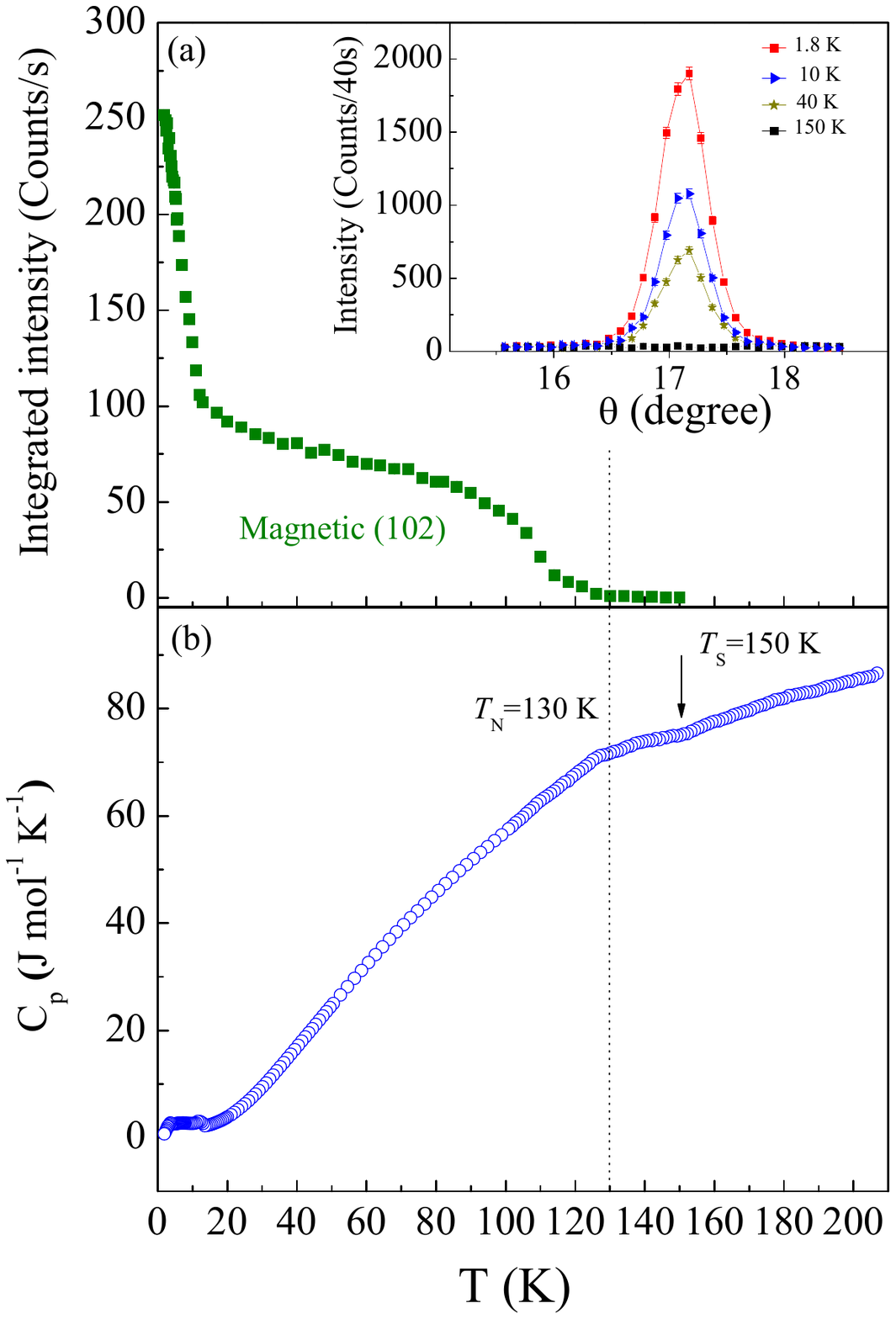}
\caption{(color online)  Temperature dependence of (a) the neutron integrated intensity of rocking-curve scan of the (102) magnetic Bragg reflection  and (b) heat capacity C$_{p}$. The inset of (a) shows the neutron diffraction rocking scans (raw data) through the magnetic Bragg peak (102) at representative temperatures.}
\label{fig:OPandSH} 
\end{figure}

    Figure \ \ref{fig:OPandSH} shows the temperature dependence of the integrated intensity of
 magnetic Bragg reflection (102) and heat capacity of the same crystal. The anomaly at $T_\texttt{S}\approx$150 K in the heat capacity measurement marks the structural transition 
from tetragonal $P4/nmm$ to orthorhombic \emph{Cmma}, consistent
with $T_\texttt{S}$ observed in previous high-resolution x-ray scattering of a crystal from the same batch growth.\cite{Li2010} Below $T_\texttt{N}\approx$130 K, the integrated intensity of
 the (102) magnetic Bragg reflection emerges and increases smoothly, evidence for the onset of the well-known stripe-like AFM ordering associated with the Fe ions (also referred to as spin-density-wave, SDW)\cite{Zhao2008,Jesche2010,McGuire2009}. Decreasing the temperature below $\approx$ 12 K leads to gradual increase of the integrated intensity of magnetic (102), suggesting the appearance of another magnetic transition. 

Using the integrated intensities from rocking curves of nuclear and magnetic Bragg reflections, we calculate the average magnetic moment $\langle gs\rangle$ 
associated with the Fe ordering, the detailed procedure of which has been described elsewhere\cite{Lee2010, Li2010}. The magnetic moment and form factor can be obtained by
          \begin{equation}
  gSf_{\rm M}(\lvert\textbf{q}\rvert)=\frac{2}{\gamma_{n}r_{e}} \sqrt{\frac{\textit{I}_{\rm M}\rm sin2\theta}{C_{\rm M}(\textbf{q})\lvert F_{\rm M}(\textit{hkl}) \rvert^{2} \rm sin^{2}\alpha})}                          \label{eq2}
     \end{equation}
      
     where $\gamma_{n}$=-1.913, $r_{e}$=2.81794$\times10^{-5}${\AA} is the classical electron radius, C$_{\rm M}(\textbf{q})$ function can be obtained 
from analyzing the data of nuclear Bragg reflections, \textit{f}$_{\rm M}$($\lvert\textbf{q}\rvert$) is the magnetic form factor at the magnetic reciprocal lattice ($\textbf{q}\textit{}$), and $\rm sin^{2}\alpha$ =1-($\hat{\textbf{q}}\cdot\hat{\mu}$)$^{2}$, where $\hat{\textbf{q}}$ and $\hat{\mu}$ are the unit vectors along the scattering vector and the direction of the Fe magnetic moment.  
$\lvert F_{\rm M}(\textit{hkl}) \rvert=\lvert\Sigma \rm{sgn}_{j} \exp[2\pi i(hx_{j}+ky_{j}+lz_{j})]\rvert$, where ($x_{j},y_{j},z_{j}$) 
represents fractional coordinates of the\textit{ j}th atom in the AFM unit cell (sgn$_{j}=\pm$). 
  
      Assuming that the Fe moments point along \textit{a} axis as reported previously \cite{Zhao2008, Maeter2009,Jesche2010, Chen2008,Jesche2009}, the observed magnetic Bragg reflections, their $q$ values, integrated intensities, and average magnetic moments are listed in Table\  \ref{tab:MagMoment}. The calculated moments were also 
corrected on the basis of the fact that the rocking curves of the nuclear Bragg peaks in our measurements include contributions from the (\textit{h} 0 0) and (0 \textit{k} 0) orthorhombic twinned domains, whereas 
the magnetic Bragg peaks are due to the (\textit{h} 0 0) domain only.\cite{Lee2010}
To estimate the average magnetic moment, we use the Fe$^{2+}$ form factor determined for SrFe$_2$As$_2$ \cite{Lee2010} and LaFeAsO\cite{Li2010}. 
\begin{figure} \centering \includegraphics [width = 1\linewidth] {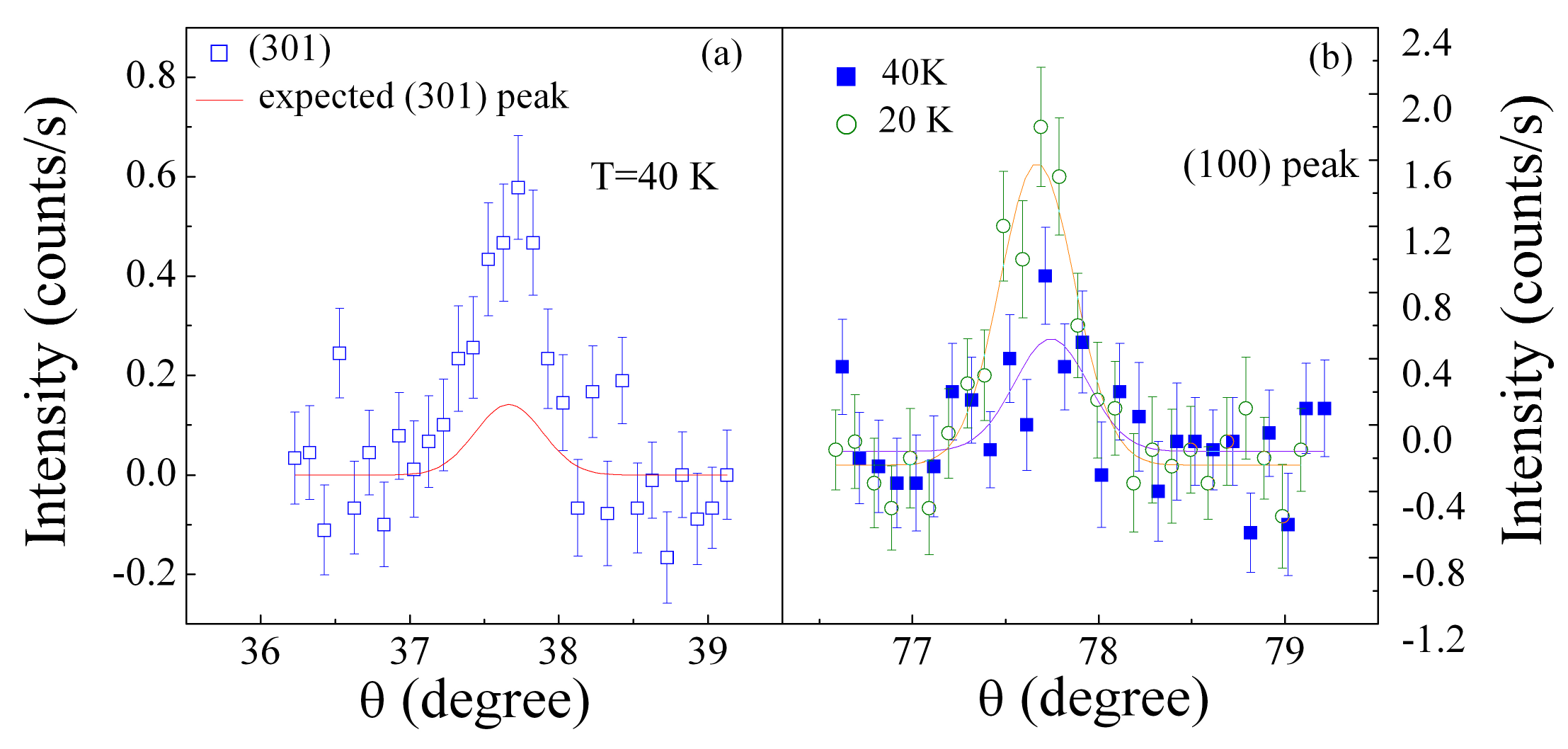}
\caption{(color online) Neutron diffraction rocking curves through the magnetic Bragg reflections of (a) (301) at 40 K and (b) 
(100) at 40 and 20 K after subtracting the background at 190 K. The solid line (a) shows the expected curve of (301) peak assuming Fe moments are along \textit{a} axis with an average magnetic moment $\langle gs \rangle$ = 0.90 $\mu_B$. The solid lines of (b) show the 
fitting curves using the Gaussian function. 
\label{fig:301-peak} }
\end{figure}

\begin{table}
\caption{The q values, experimental integrated intensities I$_{\rm exp}$, obtained average magnetic moments without Fe spin reorientation ($\omega$=0), the simulated integrated intensities 
without Fe spin reorientation I$_{\rm sim}$ ($\omega$=0) and with a Fe rotation angle ($\omega$=10$^{\rm o}$) of the magnetic reflections in consideration to an average magnetic moment $\langle gs \rangle$ = 0.90 $\mu_B$ at 40 K in CeFeAsO.} 
 \label{tab:MagMoment}
\begin{ruledtabular}
\begin{tabular} {llllll}
(\textit{h}\textit{k}\textit{l})&$q$({\AA}$^{-1}$)&I$_{\rm exp}$&$\langle gs \rangle$($\mu_B$)($\omega$=0)& I$_{\rm sim}$($\omega$=0) &  I$_{\rm sim}$($\omega$=10$^{\rm o})$ \\
  &&&& \\
\hline
(100)& 1.114   &     2.70$\pm$1.2   &         &   0  &   9.55   \\
(101)& 1.334   &    36.2$\pm$1.37  &    0.636$\pm$0.014 &  72.3 & 77.3\\
(102)& 1.844  &      80.6$\pm$1.78 &   0.905$\pm$0.014  &  81.2 & 82.6\\
(120)& 2.494  &     50.04$\pm$7.43 &  0.95$\pm$0.07 &  45.9 & 52.7 \\
(300)& 3.342   &      0     & &  0 & 0.57     \\
(301)& 3.422   &     2.97$\pm$0.52 &   1.754$\pm$0.156  & 0.78 & 1.5  \\
(303)& 4.003  &      2.19$\pm$0.45  & 0.88$\pm$0.085 &  2.34 & 2.5\\
\end{tabular}
\end{ruledtabular}
\end{table}

It can be seen in Table\ \ref{tab:MagMoment} that the average magnetic moment $\langle gs \rangle$ extracted from the (101)
 and especially (301) magnetic Bragg peaks deviates from the values extracted from other magnetic reflections. In addition, 
we observe a very weak (100) peak (see Fig.\ \ref{fig:301-peak} (b)) that is not allowed by the stripe AFM magnetic structure with the magnetic moment alone the
 $a$-axis.   
 Here, we consider a few possible explanations to this anomaly: 1) The magnetic form factor deviates from that of Fe$^{2+}$ 
due to the anisotropic $d$-orbital distributions. Based on band structure calculations, Lee et al. have argued that strong
 hybridization of Fe-As orbitals can modify the form factor in an irregular form for different reflections.\cite{Lee2010}    
 2) The Ce magnetic short range ordering due to induced Ce spins by the Fe sublattice may add anomalous contribution to 
specific peak intensities that can affect the calculation of the average Fe magnetic moments. Muon spin relaxation ($\mu$SR)
 measurements show the magnetic contribution from Ce moments is finite ($\sim$ 0.075 $\mu_{B}$ at 40 K) 
and may induce such changes to the intensities of some Bragg reflections.\cite{Maeter2009} This scenario is probably not 
valid here since all line widths of magnetic reflections, including the (101) and (301), and the weak (100) are resolution
 limited as all other magnetic reflections, such as (102) (see Fig.\  \ref{fig:OPandSH}). 
3)The Fe$^{2+}$ magnetic moments are not aligned along the \textit{a}-axis but are uniformlly rotated in the plane, 
which is the most likely explanation. We conducted such simulations and found that a small Fe spin rotation of
 $\approx$ 10$^{\rm o}$ in the \textit{ab} plane reduces the moment at (301) significantly while affecting the 
other reflections much less. This rotation with $\langle gs \rangle \approx$ 0.9 $\mu_B$ better fits the intensity of the (301) 
and also the observation of the weak (100) peak (see Fig.\ \ref{fig:301-peak}(b)). 
As shown in Table \ \ref{tab:MagMoment}, the Fe spin reorientation 
in the \textit{ab} plane increases the simulated integrated intensity to reconcile the observed weak intensity of the 
(301) reflection (see Fig. 2 (a)) and also predicts finite intensities at the (100). Our calculated Fe ordered moment 
of $\approx$ 0.9 $\mu_B$ is comparable to that determined by powder neutron diffraction, 0.8(1) $\mu_{B}$ \cite{Zhao2008}, 
but larger than the 0.34 $\mu_{B}$ inferred from the M\"{o}ssbauer spectra \cite{McGuire2009}. 

 \subsection{Two magnetic anomalies observed by neutron diffraction, heat capacity and resistivity measurements}
\begin{figure} \centering \includegraphics [width = 1\linewidth] {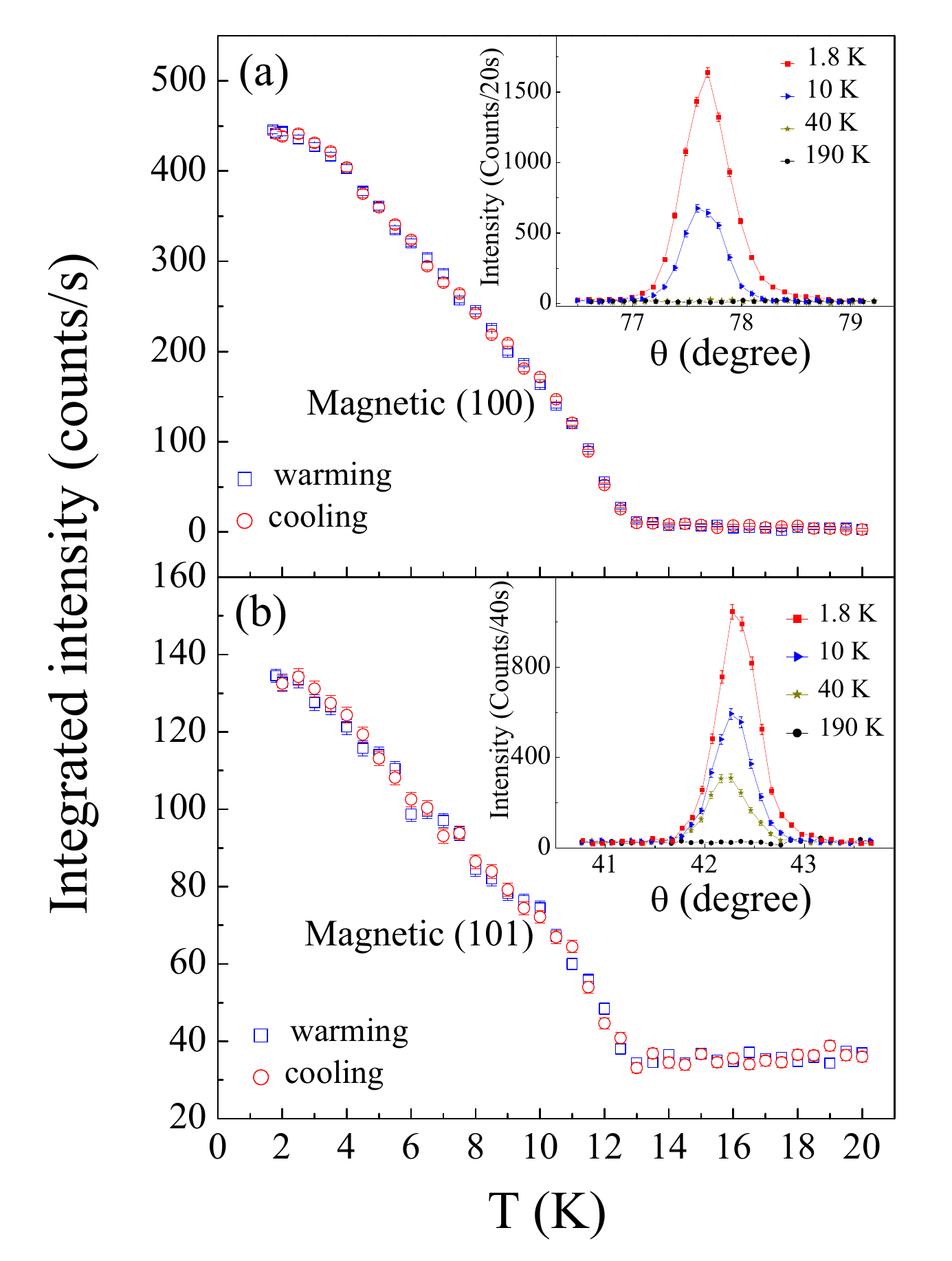}
\caption{(color online) Temperature dependence of the neutron integrated intensity for magnetic Bragg peak at (a) (100) and (b) (101). The insets show the corresponding neutron diffraction 
rocking scans (raw data) at representative temperatures. }
\label{fig:100-peak} 
\end{figure}
\begin{figure} \centering \includegraphics [width = 1\linewidth] {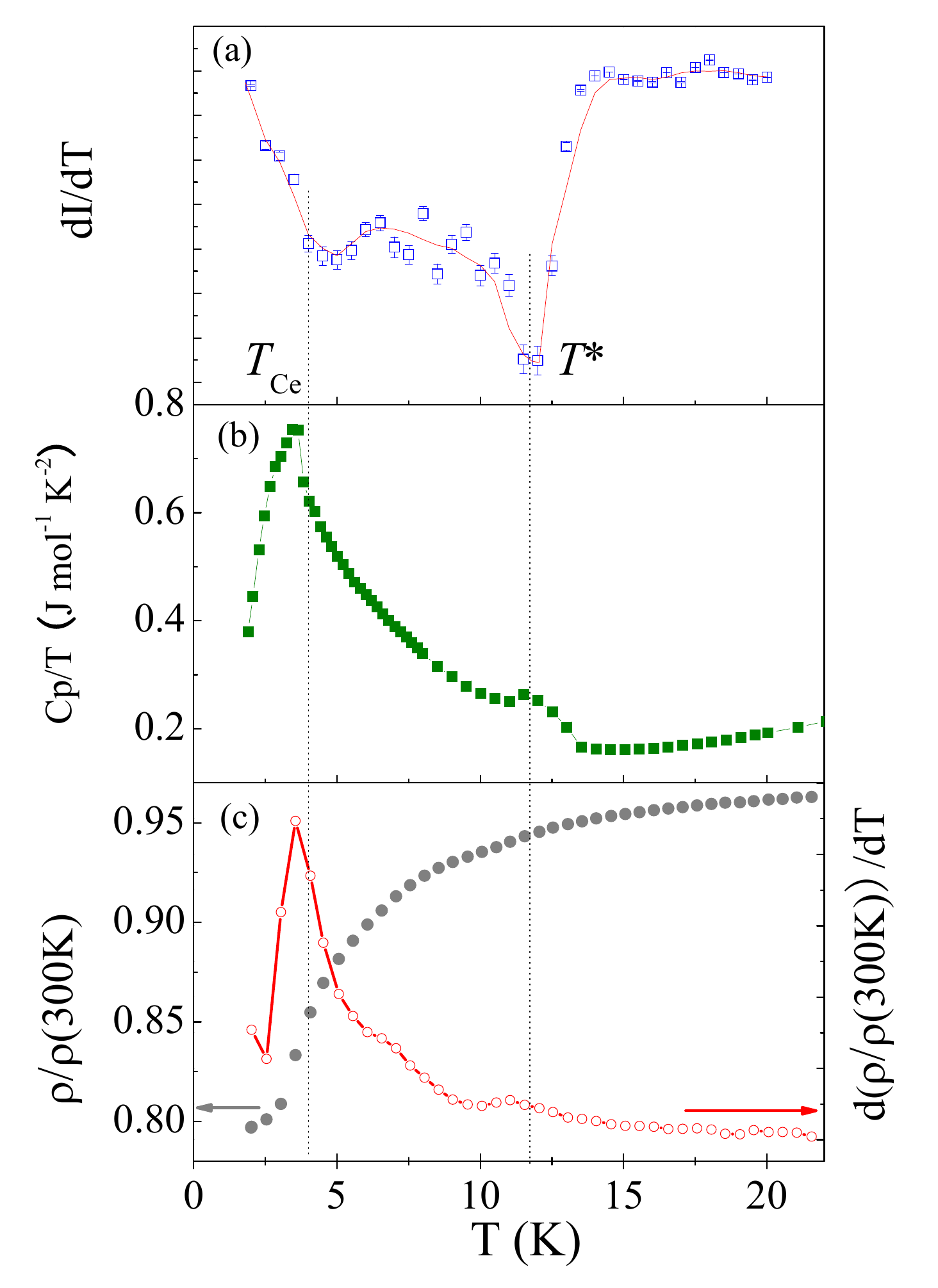}
\caption{(color online)  Temperature dependence of (a) the first derivative of the neutron integrated intensity for magnetic Bragg peak at (100) (performed
numerically), (b) the
 heat capacity $C_{p}$/$T$ at low temperatures, and (c) the normalized resistivity $\rho/\rho$(300K) (left) and its first
derivative (right).}
\label{fig:OD-derivative} 

\end{figure}

The gradual increase in the intensity of the (102) magnetic reflection at around 12 K (see Fig.\ \ref{fig:OPandSH}(a)) is accompanied with the significant increase in the intensity of (100), (300), (101) and (120) magnetic reflections. Fig.\ \ref{fig:100-peak} (a) and (b) show a detailed integrated 
intensity versus temperature of the magnetic (100) and (101) reflections, respectively, both exhibiting a weak slope change at $\approx$ 4 K in the integrated intensity \textit{versus} $T$.  The anomaly at $\approx4$ K is more obvious in the first derivative of the integrated intensity with respect to temperature (performed numerically, see Fig.\  \ref{fig:OD-derivative} (a)). The successive magnetic transitions at $\approx$ 12 K and 4 K observed in the neutron diffraction measurements can be clearly identified as the two anomalies in the heat capacity and resistivity measurements, as shown in Fig.\ \ref{fig:OD-derivative}(b) and (c).  The absence of thermal hysteresis in the integrated intensities of the magnetic (100) and (101) reflections
 upon warming and cooling the sample suggests these are second-order phase transitions. It is worthwhile noting that we also measured the heat capacity of another two pieces of CeFeAsO crystals (the 2nd piece is from the same batch and the 3rd piece from a different
batch). All samples show two anomalies at $T\approx$ 12 and 4 K although the anomaly measured on the 2nd piece at $\approx$ 12 K is slightly weaker.

\begin{figure} \centering \includegraphics [width = 1\linewidth] {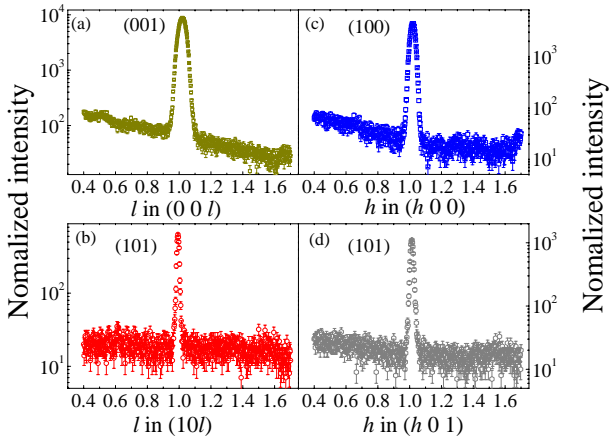}
\caption{(color online) Extended neutron diffraction scans along \textit{l} through (a) (001) and (b) (101) reflections, and along \textit{h}  through (c) (100) and (d) (101) reflections at 1.8 K. }
\label{fig:long} 
\end{figure}

      Note that our neutron diffraction result on (102) reflection is different from a previous report on polycrystalline CeFeAsO, where the (102) magnetic Bragg reflection shows initial
 increase at $T \approx20$ K and a much more abrupt increase at $T\approx 4 $ K.\cite{Zhao2008}  In addition, in this polycrystalline study it was also reported that (0 0 $\frac{1}{2}$) peak was observed and associated with the ordering of the Ce. In contrast, our single crystal study shows no detectable peak at the (0 0 $\frac{1}{2}$) as shown in Fig.\ \ref{fig:long}. In fact, long $h$ and $l$ scans along principal directions at $T = 1.8$ K do not show any evidence of magnetic unit cells that extend beyond the chemical cell, as shown in Fig.\  \ref{fig:long}. Qualitatively, these observations are evidence that the repeat unit of all the magnetic orderings are confined to the chemical unit cell. But, the results also raise the question about the two anomalies observed in the specific heat, resistivity and in the order parameter measured on the (100), (101), (120) or (102) reflections on CeFeAsO crystals. To resolve these questions and get insight into the
 nature of the two magnetic transitions, we conducted resonant x-ray magnetic scattering (RXMS) study on CeFeAsO taking advantage that this technique is element specific. 


\subsection{Ce $L_{\rm II} $ RXMS measurements}
\begin{figure} \centering \includegraphics [width = 1\linewidth] {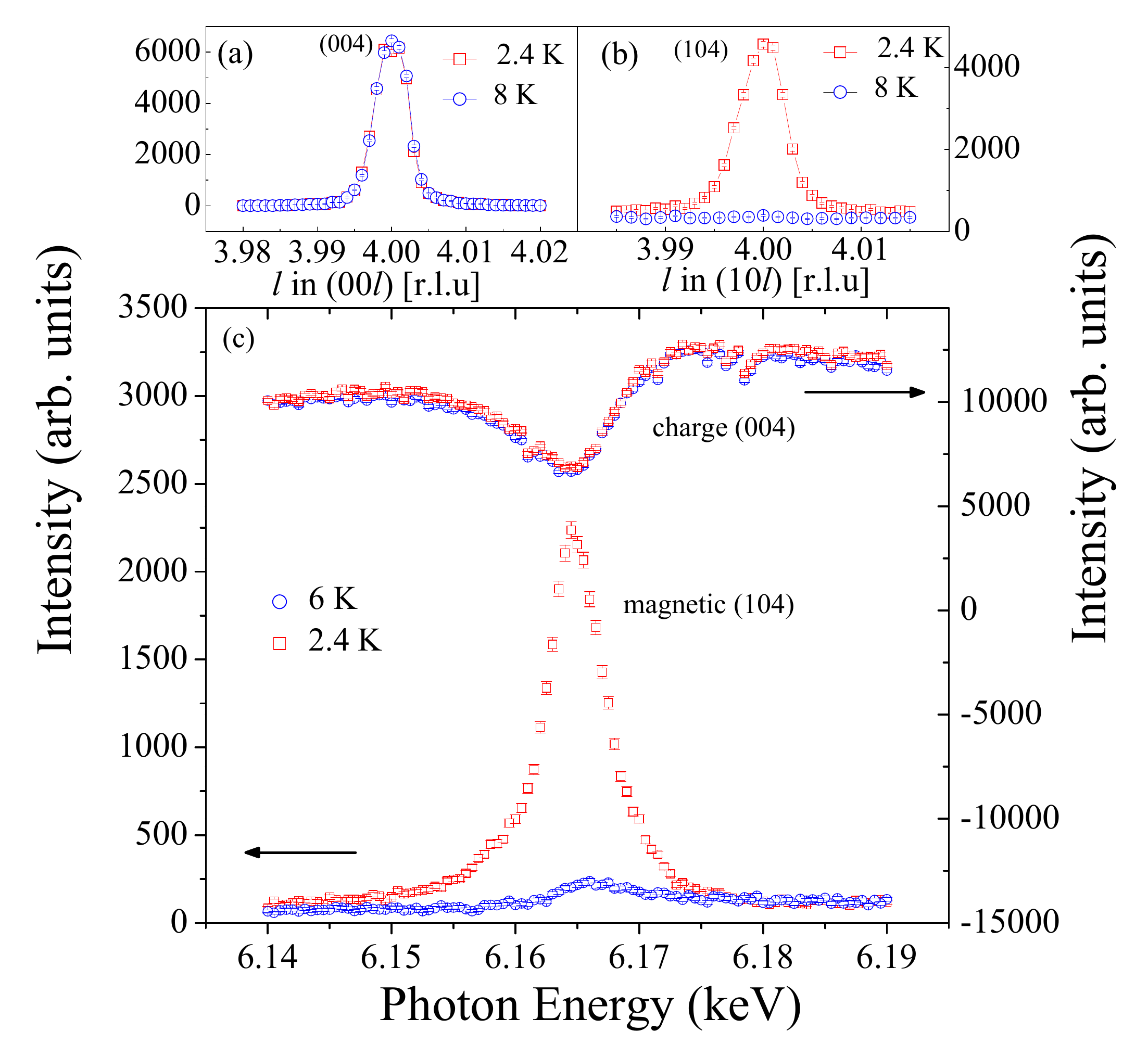}
\caption{(color online)  Resonant x-ray magnetic scattering at the Ce-$L_{\rm II}$ edge \textit{L} scans through 
(a) charge reflection (004) Bragg reflection and (b) magnetic reflection (104) at 2.4 and 8 K. (c) Photon energy 
scan at constant-Q values through the Ce $L_{\rm II}$ resonance  at $T = 2.4$ and 6 K for the 
strictly charge (004) Bragg reflection and the (104) magnetic Bragg reflection.}
\label{fig:RXMS1} 
\end{figure}
\begin{figure} \centering \includegraphics [width = 1.\linewidth] {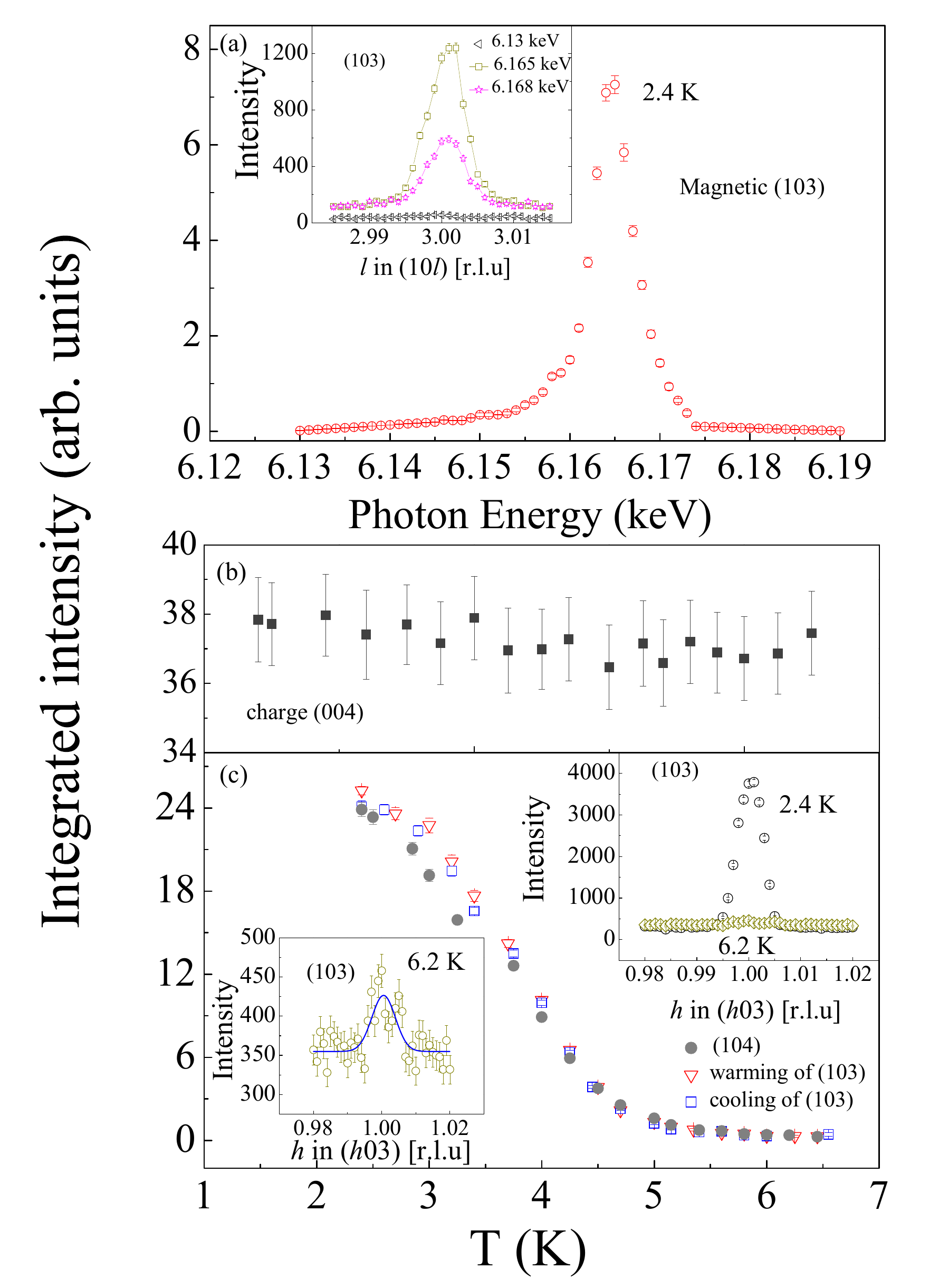}
\caption{(color online) (a) Energy dependence of resonant magnetic scattering integrated intensities collected at $T$ = 2.4 K through the Ce $L_{\rm II}$ edge for fixed $Q$ at (103) magnetic Bragg reflection. The 
inset shows the representative distribution of \textit{l} scans through the magnetic peak (103) measured at different energies at 2.4 K. (b) Temperature dependence of the integrated intensity of charge reflection (004) and (c) magnetic reflections (103) and (104) at E=6.165 keV. The right inset of (c) shows the resonant x-ray magnetic scattering \textit{h} scans through magnetic reflection (103) at 2.4 and 6.2 K. The left inset of (c) is the enlarged resonant x-ray magnetic scattering \textit{h} scans through magnetic reflection (103) at 6.2 K}
\label{fig:RXMS2} 
\end{figure} 
Figures\ \ref{fig:RXMS1} (a) and (b) show \textit{L} scans through the chemically allowed charge peak (004) and 
a newly emerging peak (104) at $T= 8$ and 2.4 K, 
measured at x-ray photon energy $ E = 6.165$ keV, close to the Ce $L_{\rm II}$ edge. Whereas the (004) is practically 
unchanged at the two temperatures, the (104) emerges at 2.4 and 
disappears above $\approx$ 7.4 K. To confirm the magnetic origin of the (104) peak, we conducted a photon energy scan at constant $Q$ value through the Ce $L_{\rm II}$ edge at (104) and a similar scan at the (004) charge reflection at 2.4 and 6 K (see Fig.\ \ref{fig:RXMS1}(c)). At 2.4 K, a clear resonance peak at 6.165 keV is observed for the (104) Bragg reflection in the $\sigma-\pi$ scattering geometry where most of the charge resonant scattering signal appearing mainly in the $\sigma$-$\sigma$ geometry is suppressed, which is typical to resonant magnetic scattering at the $L_{\rm II}$ edge of rare earth containing compounds\cite{Kim2005}. We also note that the resonance energy is shifted by approximately 1 eV above the Ce $L_{\rm II}$ absorption edge, as determined from the inflection point of the charge (004) signal. Such a shift is evidence that the resonance originates from dipole transitions between the core 2\textit{p} and the partially filled 5\textit{d} states\cite{Nandi2009,Hannon1988}. Similar resonant magnetic scattering is observed in energy scan through the Ce $L_{\rm II}$ edge for fixed $Q$ at (103) reflection (not shown here) and also in the energy dependence of the integrated intensity of (103) peak at 2.4 K as shown in the Fig. 7(a), confirming that both (103) and (104) are Ce-specific magnetic reflections. 
     
     The temperature dependence of the (004) charge Bragg reflection as well as the (103) and (104) magnetic reflections are shown in Fig.\ \ref{fig:RXMS2} (b) and (c), respectively. 
Whereas the (004) reflection is practically constant as a function of temperature (within uncertainties), the (103) and (104) reflections are strongly temperature dependent and show
 a transition at 4 K, corresponding to the peak in the derivative of the integrated intensity for both (103)
and (104) magnetic reflections with respect to temperature. The gradual increase in the integrated intensity and the absence of the thermal hysteresis upon both warming and cooling processes 
indicate the second-order nature of the transition. All these results confirm the formation of Ce AFM long-range ordering \textit{via } Ce-Ce exchange interaction below $T_\texttt{Ce}\approx$ 4 K. 
Note that this is consistent with magnetic susceptibility measurements, which show an effective moment of 2.25 $\mu_{B}$ per Ce$^{3+}$ ion, close to the free-ion value of 2.54 $\mu_{B}$ and a negative Weiss temperature of $\approx$ -17 K indicative 
of a dominant AFM Ce-Ce coupling.\cite{McGuire2009} The measured transition temperature by RXMS, $\emph{T}_\texttt{Ce} \approx 4 $ K, is in good agreement with the anomaly observed in our heat capacity, resistivity measurements and also the low temperature anomaly observed in the (100), (101), (120) and (102) neutron diffraction studies. 
       
        We point out that a tail in the resonant x-ray magnetic scattering \textit{Q} scans for magnetic (103) and (104), i.e., nonzero integrated intensity is observed above \emph{T}$_\texttt{Ce}$=4 K up to at least 7.4 K. As shown in the insets of Fig. 7 (c) at a representative temperature of 6.2 K, the linewidth of the magnetic (103) peak is approximately 50\%  broader than that at 2.4 K. We estimate that the Ce magnetic correlation length at 6.2 K
is shorter than 150 \AA. This short-range Ce ordering is most probably due to the polarized Ce moments by Fe-Ce coupling as observed by $\mu$SR measurements on polycrystalline CeFeAsO sample\cite{Maeter2009}. 

\subsection{Nature of the transition at $T^*$, Ce-Fe coupling and Ce long-range AFM magnetic structure}
Whereas the RXMS shows a single transition associated with the Ce ordering at $T_\texttt{Ce}$ $\approx 4$ K, the neutrons show abrupt increase in the intensity of the magnetic reflections, such as (100), (101), (102) and (120) at $T^{*} \approx 12$ K, and only a subtle anomaly at $T_\texttt{Ce}$. 
 We emphasize that the neutron and RXMS were conducted on the same crystal. Moreover, we also performed RXMS measurements on the third piece of crystal from a different batch and obtained 
 reproducible results. Since either Fe spin reorientation, short range ordering development of Ce, or both may be responsible for the anomaly at $T^*$, we further examine the nature of this anomaly. Below we argue that the anomaly is mainly due to Fe 
spin reorientation, with a probable minor contribution from Ce short-range ordering. 

\begin{figure} \centering \includegraphics [width = 1.\linewidth] {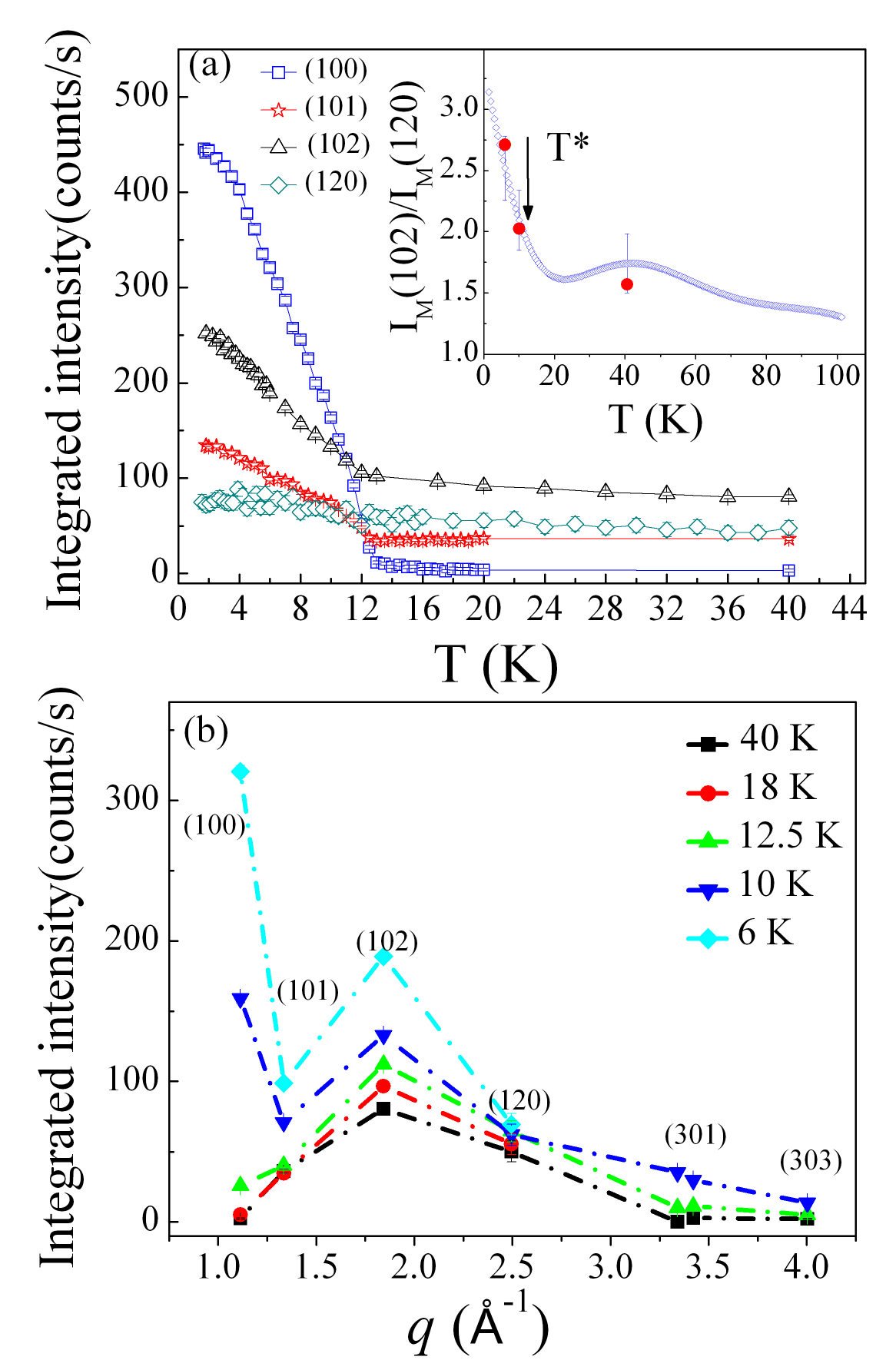}
\caption{(color online) (a) The temperature dependence of the integrated intensities of several magnetic reflections. In the inset, the blue open symbols with three representative error bars show the ratio between the experimental integrated intensities of (102)
and (120) magnetic reflections. The red solid symbols show the simulated results on this ratio at 40, 10 and 6 K under the absolute values 
of $\omega$ of 10$^{\rm o}$, 45$^{\rm o}$ and 60$^{\rm o}$, respectively. (b) The \textit{q} dependence of the experimental integrated intensities for the magnetic reflections at different temperatures.}
\label{fig:experimentaldata} 
\end{figure}

     As shown in Fig.\ \ref{fig:experimentaldata}(a), the (100) integrated intensity, not allowed by the Fe SDW magnetic structure with the moment along \textit{a}-axis, is negligible compared to other magnetic reflections such as (101), (102), (120) above $T^*$, but the (100) integrated intensity increases significantly and becomes the strongest peak below $\approx$ $T^*$. This special phenonmenon thereby leads to two distinct 
features below/above $T^*$ in the \textit{q} dependence of the integrated intensity of the different magnetic reflections in Fig.\ \ref{fig:experimentaldata} (b). 
Another important observation in our experiment is that the temperature dependence of the intensity ratio of (102)/(120) exhibits an increase below $T^*$, as shown in the inset of Fig.\ \ref{fig:experimentaldata}(a). If there is no Fe spin reorientation, the short-range Ce ordering by itself would have yielded a constant in the temperature depedence of this ratio. 
Moreover, as shown in Fig. (1) and (3), we notice that the full-width at half maximum (FWHM) in the rocking curve of the new emerged (100) magnetic peak at 10 K is very similar to the almost temperature-independent FWHM values of long-range (101) and (102) magnetic Bragg reflections and also nuclear Bragg reflections below $T_\texttt{N}$. This is evidence that (100) peak is due to long-range ordering and therefore is not induced by the short-range  Ce ordering.  
 
      The appearance and a significant increase in intensity of the long-ranged (100) magnetic Bragg peak and the anomalous intensity ratio of the (102)/(120) can be explained by 
a uniform rotation of the Fe moment in the $ab$-plane away from the $a$-axis as expressed by the $\sin^2\alpha$ term in Eq. (1) where $\alpha$ is the angle between a 
scattering vector and the spin direction. The angular dependence of the magnetic structure factors ($\mid$F$_{\rm M}$(hkl)$\mid$)$^2$sin$^{2}\alpha$ in the \textit{ab}-plane
 due to this term is shown in Fig.9(a). It can be seen 
that with the increase of angle $\omega$ between Fe moment and \textit{a}-axis in the \textit{ab} plane, the magnetic structure factor of (100) peak increases much faster than (101) and (102) peaks, which has a tendency to result in the highest integrated intensity.  
We simulated the magnetic integrated intensity $I_{\rm M}$ at 10 K and 6 K by considering various $\omega$ values and adjusting the average
magnetic moment $\langle gS \rangle$. As shown in Fig. 9(b), the Fe spin reorientation with absolute value of $\omega$ of $\approx$ 45$^{o}$, together with the increase of magnetic moment to be $\approx$ 0.98 $\mu_{B}$ is consistent with the experimental result at 10 K. The simulated intensity ratio of I$_{\rm M}$(102)/I$_{\rm M}$(120) is also consistent with the experimental result, as shown in the inset of Fig. 8 (a). Alternatively, calculation of g\textit{S}\textit{f}$_{\rm M}$($\lvert\textbf{q}\rvert$) according to Eq. 1 using the experimental data at 10 K, shows an irregular and unphysical form
factor if there is no Fe spin reorientation. This can be rectified by a rotation of the Fe spins to about 45 degrees in \textit{ab }plane that shifts the
 g\textit{S}\textit{f}$_{\rm M}$($\lvert\textbf{q}\rvert$) values significantly to yield a smoother Fe$^{2+}$ form factor, with an average magnetic 0.98 $\mu_B$, as shown 
in the inset of Fig. 9 (b). It is interesting to point out that with the increase of 
the absolute values of $\omega$, there is a crossover between two distinct features in the \textit{q} dependence of simulated $\textit{I}_{\rm M}$, showing  similar behavior to the experimental $\textit{I}_{\rm M}$ with decreasing the temperatures (see Fig.\ \ref{fig:experimentaldata} (b)).
This strongly suggests that at $T^*$ significant uniform Fe spin reorientation away from \textit{a} axis in the \textit{ab} plane takes place. This is further confirmed by our simulation at 6 K that shows
 a Fe spin reorientation angle of $\approx$ 60 degree with an average magnetic moment of 1.1 $\mu_{B}$. Moreover, the simulated intensity ratio of I$_{\rm M}$(102)/I$_{\rm M}$(120) is also within the error bar of the experimental intensity ratio at 6 K (see Fig. 8 (a)). We also conducted similar simulations  assuming Fe spin reorientation in the \textit{ac} plane, and found it is inconsistent with the experimental data at 10 K and 6 K. Therefore, the anomaly at $T^*$ in CeFeAsO is mainly ascribed to the Fe spin reorientation away from a-axis in the \textit{ab} plane while preserving the underlying SDW magnetic structure. The overlap 
of Ce magnetic Bragg peaks with those of reoriented Fe spins makes it impossible to state whether Fe spins continue to reorient below $T_\texttt{Ce}$. 
\begin{figure} \centering \includegraphics [width = 1.\linewidth] {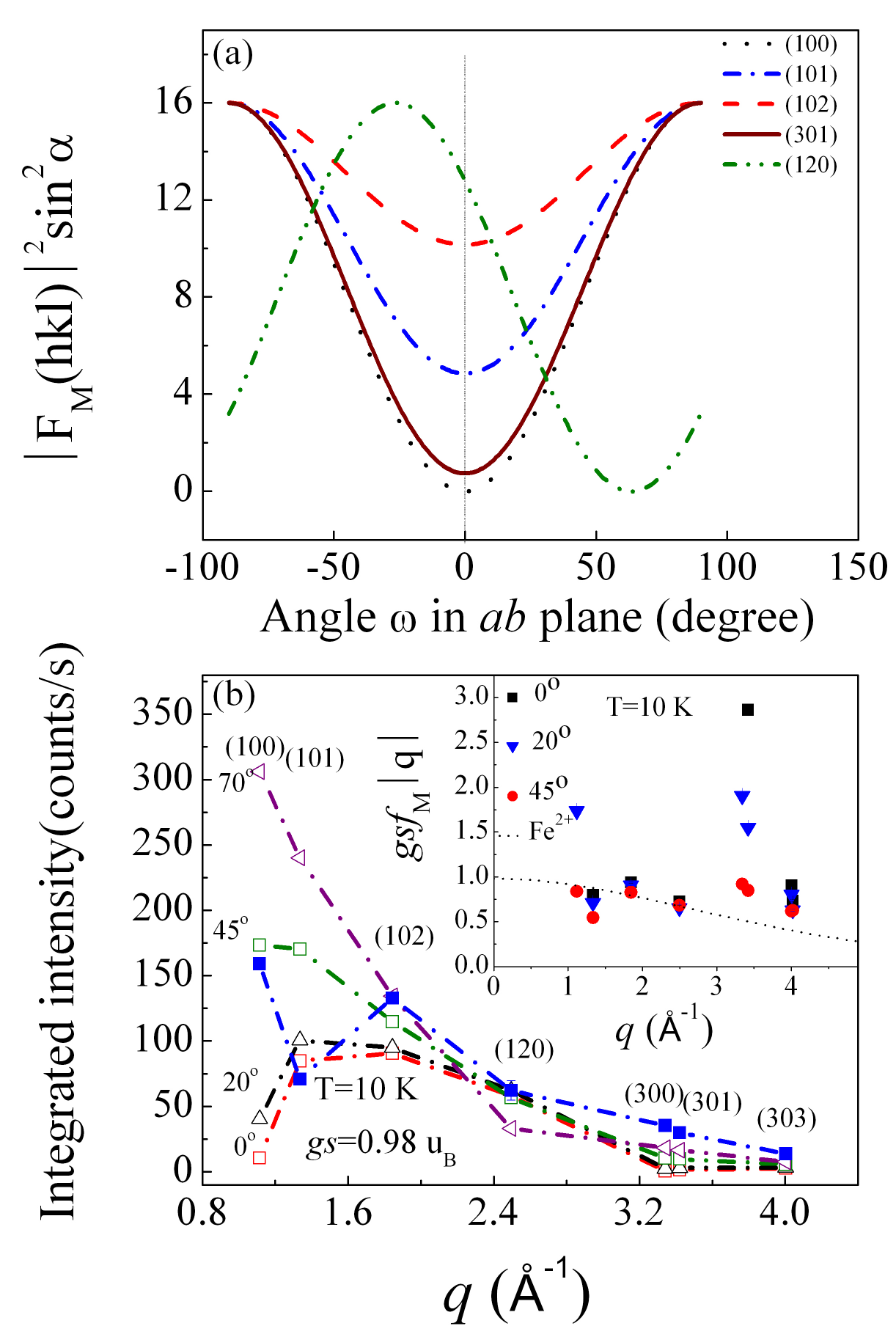}
\caption{(color online) (a) The magnetic structural facotors ($\mid$F$_{\rm M}$(hkl)$\mid$)$^2$sin$^{2}\alpha$ for the different magnetic reflections as a function of the angle $\omega$ between \textit{a} axis and the direction of Fe magnetic moment within \textit{ab} plane. (b) The 
simulation (open symbols) on the \textit{q} dependence of integrated intensities for the magnetic reflections under different absolute values of $\omega$ at 10 K. 
The solid squares show the experimental data at 10 K. The inset illustrates the simulation on the \textit{q} dependence of the gS\textit{f}$_{\rm M}(\lvert\textbf{q}\rvert)$ under representative absolute values of $\omega$ of 0, 20 and 45 degree.}
\label{fig:simulations} 
\end{figure}     

    The Fe spin reorientation proposed is not unique to for CeFeAsO. It was observed in the iso-structural PrFeAsO where the in-plane stripe-like Fe AFM order is preserved but the Fe moments rotate to out-of-plane collinear AFM order below $T_{\rm Pr}$ $\approx$ 11 K via Pr-Fe interaction.\cite{Maeter2009,McGuire2009}  
Note that whereas the long-range Pr moments point along the $c$-axis and the Fe moments rotate toward that axis in PrFeAsO, the long-range Ce moments are in the \textit{ab} plane as 
discussed below and the Fe moments rotate in the\textit{ ab} plane in CeFeAsO, possibly reflecting the magnetic anisotropy of Fe sublattice by Ce magnetism via the Ce-Fe interaction in CeFeAsO. 
Based on our analysis, we cannot rule out the possibility that short-range Ce ordering slightly contributes to the anomaly at $T^*$. Our RXMS results indicate gradual but less prominent Ce polarization up to at least 7.4 K as is expected from the mutual effect of the Ce-Fe coupling on both sites.
Since the anomaly at $T^*$ is mainly ascribed to the Fe spin reorientation, we hypothesize that 
 the induced finite Ce moments extends even to a higher temperature $\approx$ 120 K, as observed in $\mu$SR studies \cite{Maeter2009}. 
The mutual influence between Fe and Ce sublattice above $T_\texttt{Ce}$ 
indicates that the Ce-Fe coupling due to non-Heisenberg anisotropic exchange as previously suggested\cite{Maeter2009}, is strong in CeFeAsO. This is consistent with 
the rather large Ce-Fe coupling constant 65.3 T/$\mu_{B}$ on the order of the Ce-Ce exchange interaction obtained from the $\mu$SR measurements\cite{Maeter2009}. The strong
Ce-Fe coupling is also supported by inelastic neutron scattering \cite{Chi2008,LiS2010} experiments showing the Ce crystalline electric field split. 
      
   We point out that the mutual Ce-Fe coupling in CeFeAsO is similar to that observed in other systems involving rare-earth and transition-metal couplings, such as 
in NdFeO$_{3}$\cite{Bartolome1997}, NdNiO$_{3}$\cite{Garc1994,Bartolom2000}, DyVO$_{3}$ \cite{Zhang2012} and in NdCrO$_{3}$ \cite{Bartolome2000}. For example, 
in NdFeO$_{3}$, Nd moments are slightly influenced by iron moments below $T\approx 25$ K, and are fully ordered only at 1 K. A spin reorientation transition occurs in a
 wide temperature range from 167 to 125 K, with a gradual rotation of Fe moments in the \textit{ac} plane from \textit{G$_{x}$F$_{z}$} to \textit{G$_{z}$F$_{x}$} magnetic structure.\cite{Bartolome1997}. 
For DyVO$_{3}$, a complex temperature-induced V spin reorientation from C-G-C type AFM ordering occurs with the decrease of temperature. The ordering of Dy moment emerges at $\approx$ 18 K, whereas a slight ordering of Dy moment via Dy-V coupling
starts at $\approx$ 36 K.\cite{Zhang2012}. 
In the case of NdCrO$_{3}$, while the Cr spin reorientation happens at $\approx$34.2 K, the Nd-Cr interaction starts to polarize the Nd moments at $\approx$ 11 K before Nd becomes long-range ordered\cite{Bartolome2000}.
 \begin{figure} \centering \includegraphics [width = 1.\linewidth] {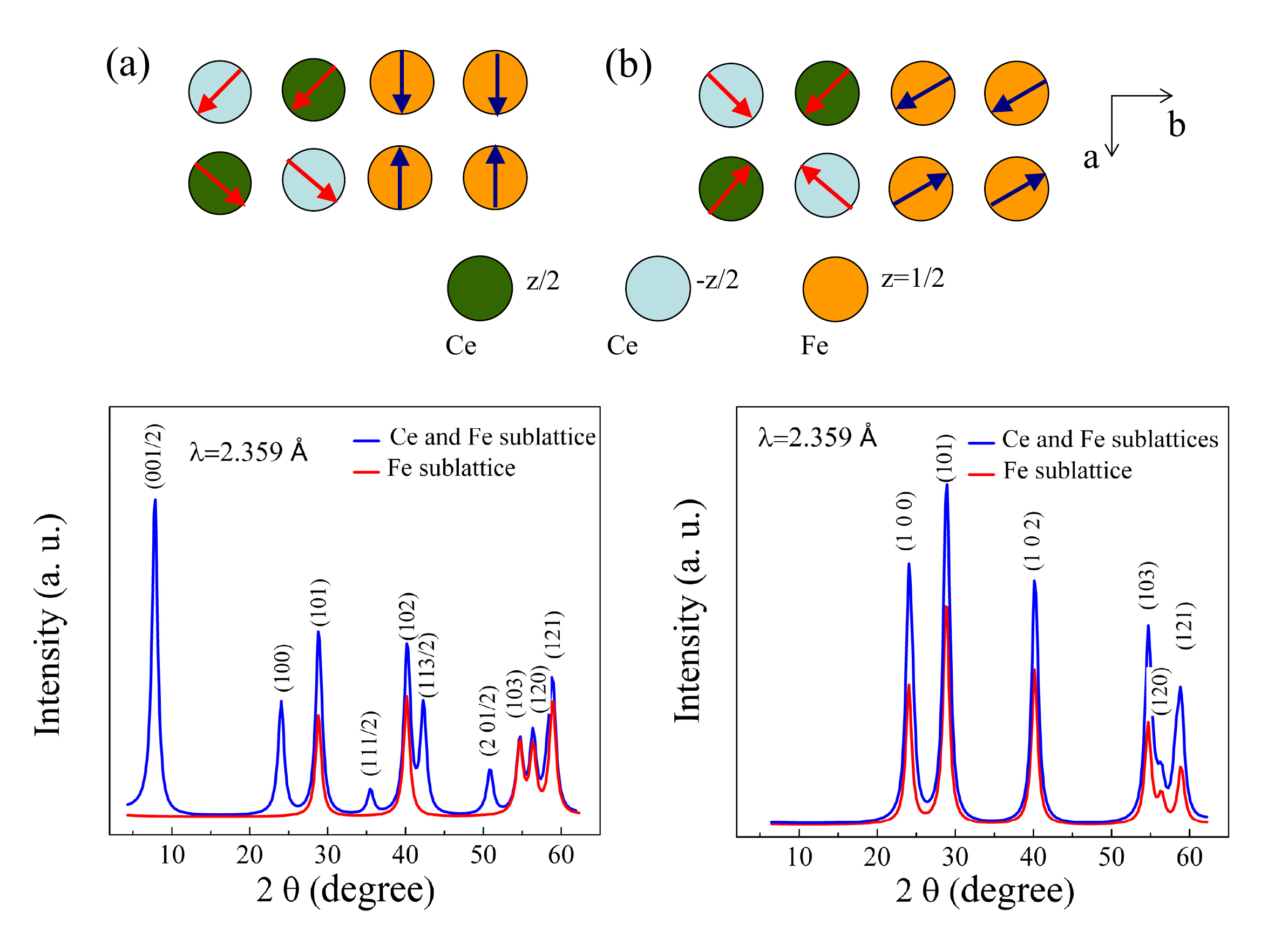}
\caption{(color online) (a) The magnetic unit cells of Ce and Fe in polycrystalline CeFeAsO below $\emph{T}_\texttt{Ce}$ proposed by Zhao \textit{et al.} \cite{Zhao2008}. The bottom figure 
shows our simulation on the 
neutron diffraction patterns based on their model. The z coordinate of Ce
 is $\approx$ 0.1413. (b) Our proposed magnetic unit cells of Ce and Fe, and the simulation 
on corresponding powder neutron diffraction patterns below $\emph{T}_\texttt{Ce}$ in single-crystal CeFeAsO.} 
\label{fig:simulations} 
\end{figure}

\begin{figure} \centering \includegraphics [width = 1.\linewidth] {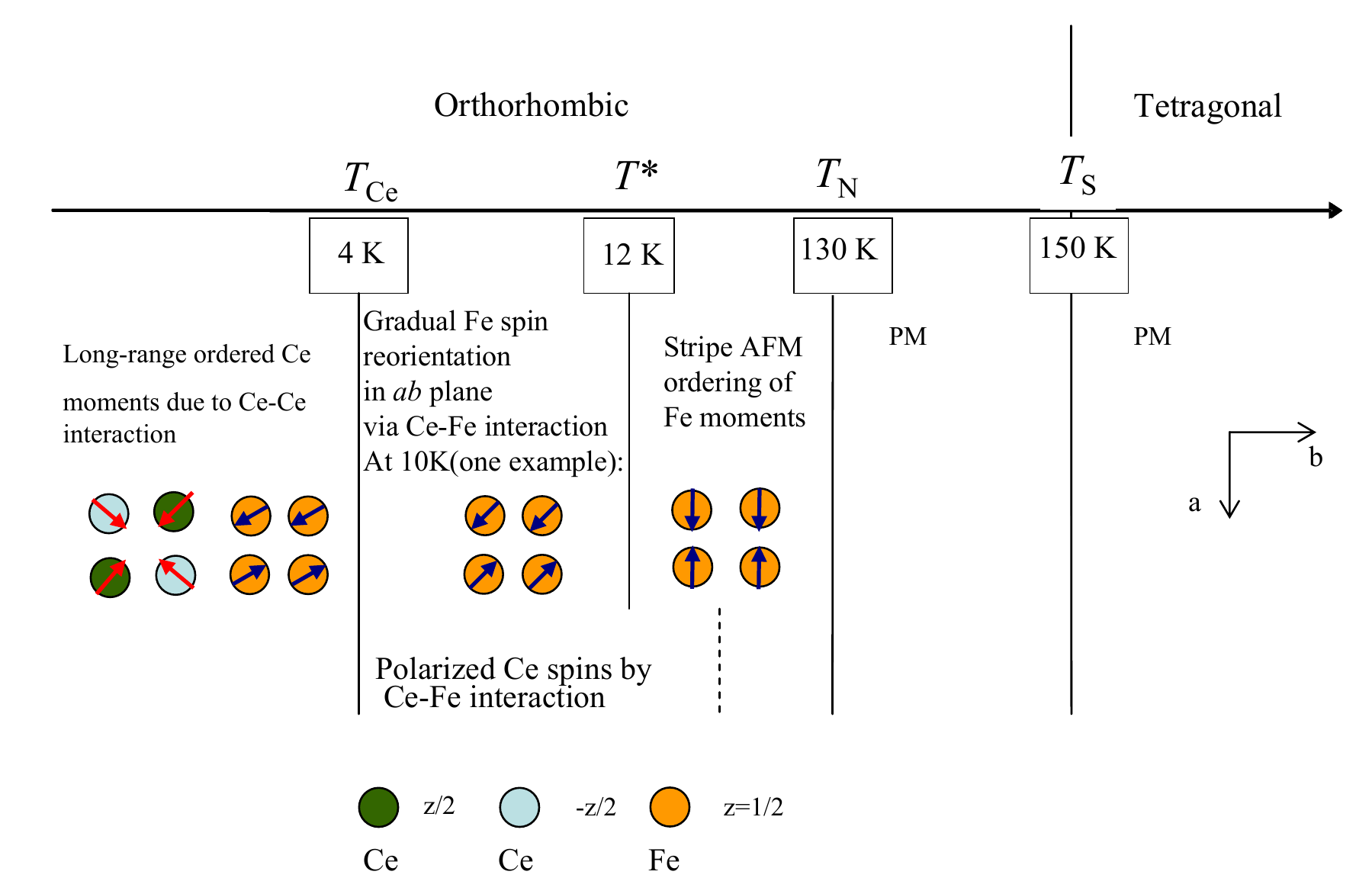}
\caption{(color online) Schematic illustration of our proposed ordering processes of structural and complex magnetic transitions
 for Ce and Fe sublattices in CeFeAsO single crystal. The dashed line indicates
 the polarized Ce spins exist up to one temperature above $T_\texttt{Ce}$. Our detailed analysis suggests possible occurrence of a small Fe spin reorientation
above $T^*$. }
\label{fig:sequence} 
\end{figure}

       We now turn to discuss the long-range Ce AFM ordering below $\emph{T}_\texttt{Ce}\approx 4$ K. Zhao \textit{et al. }\cite{Zhao2008} proposed a magnetic structure of Ce sublattice based on their powder neutron measurements. The projections of nearest-neighbor Ce moments in the \textit{ab} plane are perpendicular to each other
with slight tilting of Ce moments towards the \textit{c}-axis, forming a non-collinear magnetic structure. Based on this model, we simulated the magnetic powder diffraction due to the Fe SDW and the combined Ce-Fe ordering, as shown in Fig.\ \ref{fig:simulations}(a). This model predicts a few new magnetic reflections such as, (0 0 $\frac{1}{2}$),  (1 1  $\frac{1}{2}$), (2 0  $\frac{1}{2}$) 
below $\emph{T}_\texttt{Ce}$, whereas our single crystal neutron diffraction measurements do not show any of these reflections, especially the strongest (0 0 $\frac{1}{2}$). 
The inconsistency below $\emph{T}_\texttt{Ce}$ indicates that the long-range ordering of Ce is likely sensitive to the differences in sample form, i.e., polycrystalline \textit{versus} 
single-crystalline. Indeed, magnetic susceptibility and transport properties of three different samples (two crystals and polycrystalline CeFeAsO) were reported to exhibit 
sample dependent properties even on the Fe ordering.\cite{Jesche2010} 
As compared to the polycrystalline samples, the single crystals are believed to be more stoichiometric, especially when samples involve oxygen element.\cite {Li2010} Here, we propose another non-colinear magnetic structure of Ce sublattice based on our neutron diffraction measurements on single crystal CeFeAsO: the nearest-neighbor Ce moments
in the \textit{ab} plane are antiparallel and the other pair in the adjacent plane are perpendicular to the first one. The long-range Ce and Fe ordering below $T_\texttt{Ce}$ in CeFeAsO crystal is illustrated in Fig. 10 (b) (For lack of further evidence, the Fe spin reorientation angle 
below $\emph{T}_\texttt{Ce}$ is assumed to be close to 60$^{o}$ as obtained at 6 K). Our proposed sequence of structural and magnetic transitions for both Fe and Ce sublattices in CeFeAsO is summarized in Fig.\ \ref{fig:sequence}. 

\section{Conclusion}
The low-temperature magnetic structures, the ordered magnetic moments and the interplay between Ce and Fe magnetism in CeFeAsO single crystal are reported.
 While the Fe AFM stripe order emerges below \emph{T}$_\texttt{N}$, significant Fe spin reorientation transition occurs in the \textit{ab} plane as the temperature decreases below $T^*\approx$ 12 K. RXMS shows
that the Ce spins undergo long-range non-colinear AFM ordering below $T_\texttt{Ce}$ = 4 K.  Partial polarized Ce spin ordering likely occurs 
above $\emph{T}_\texttt{Ce}$. The temperature evolution of Ce moments and the interplay between Ce and Fe in CeFeAsO have been proposed, the picture of which is similar to other rare-earth-based NdFeO$_{3}$ NdNiO$_{3}$, DyVO$_{3}$ and NdCrO$_{3}$ systems. The effect of the strong Ce-Fe coupling on the rearrangement of Fe ordering is yet another example of the vulnerability of the Fe spin-density-wave to perturbations such as minute doping or relatively low applied pressures.

\section{Acknowledgments}
Research at Ames Laboratory is supported by the US Department of Energy, Office of Basic Energy Sciences, Division
of Materials Sciences and Engineering under Contract No. DE-AC02-07CH11358. Use of the Advanced Photon Source
at Argonne National Laboratory and the high flux isotope reactor at the Oak Ridge National Laboratory,
 was supported by the US Department of Energy, Office of Science, Office of Basic
Energy Sciences, under Contract No. DE-AC02-06CH11357. Work at ORNL was supported by the U.S. Department of Energy, 
Basic Energy Sciences, Materials Sciences and Engineering Division (JQY) and the Scientific User Facilities Division (WT and JLZ).

\end{document}